# Development of high-frequency magnetic probe for plasma diagnostics of XuanLong-50

Mingyuan WANG[1,2,*], Xiuchun LUN[1,2], Xiaokun BO[1,2], Bing LIU[1,2,3], Adi LIU[4], Yuejiang SHI[1,2,*]

[1]Hebei Key Laboratory of Compact Fusion, Langfang 065001, China

[2]ENN Science and Technology Development Co., Ltd., Langfang 065001, China

[3]Institute of Fusion Science, School of Physical Science and Technology Southwest Jiaotong University 610031 Chengdu, China

[4]Department of Plasma Physics and Fusion Engineering, University of Science and Technology of China, Anhui Hefei 230026, China

[*]E-mail of corresponding author: wmyuan@mail.ustc.edu.cn and shiyuejiang@enn.cn

## Abstract

A high-frequency magnetic probe has been designed and developed on XuanLong-50 (EXL-50) spherical torus to measure high-frequency magnetic field fluctuations caused by energetic ions and electrons in the plasma. The magnetic loop, radio filters, radio-frequency (RF) limiter, and data acquisition system of the probe are described in detail. The results of the preliminary test show that the probe can have a frequency response within the 1–200 180 MHz range. The fluctuation data from the EXL-50 plasma were analyzed in the time-frequency domain using fast Fourier transforms. Using this diagnostic system, distinct high-frequency instabilities were detected. In particular, significant frequency chirping was observed, which is consistent with the bump on tail drive instability predicted by the Berk-Breizman model.

**Keywords**:

## 1. Introduction

High-performance plasma in tokamaks requires accurate diagnosis of energetic particle populations. Key obstacles in reaching critical thermonuclear fusion temperatures include superthermal ions and electrons produced by neutral beam injection (NBI), radio-frequency (RF) waves, or fusion alphas. Energetic particles often cause instability through wave-particle resonances, which deteriorate confinement and can also result in fast-ion and electron loss that can damage the plasma-facing components [1].

Numerous fusion devices have been used to study high-frequency magnetic fluctuations caused by energetic particles (ions and electrons), including ASDEX-U [2-4], DIII-D [5-7], KSTAR [8,] JET [9, 10], JT-60U [11], LHD [12, 13], EAST [14, 15], and LH-2A [16]. NBI has been used to detect energetic ion-excited ion cyclotron emission (ICE) [5, 10, 11, 17–19]. The velocity anisotropy of runaway electrons during runaway discharge can stimulate electromagnetic radiation at a variety of frequencies [20, 21]. To study high-energy particle drive instabilities, high-frequency magnetic probes are frequently used [3, 6, 8, 21, 22].

Xuanlong-50 (EXL-50) is a medium-size spherical torus without central solenoid. At present, EXL-50 uses three sets of electron cyclotron wave (ECW) systems with 28 GHz/50 kW (330° port) and two sets of 28 GHz /500 kW (0° and 300°) to initiate start-up [23] and drive the plasma current. In a typical discharge, ECW produces a large number of energetic electrons with low electron density [24]. The electron distributions of EXL-50 may exhibit spatial gradients, anisotropy, or positive velocity gradients, all of which can cause instability via wave-particle resonances [25, 26]. The instability caused by the energetic electron is one of the most significant physical issues in EXL-50. To address this, a high-frequency magnetic diagnostic is required to assess the instability of the energetic electron drive.

A general resonance condition for the coupling of relativistic electrons to a wave may be written as: $\omega - kv - l\omega_{ce}/\gamma = 0$, where $\omega$ is the instability frequency, $k$ is the wave number, $v$ is the velocity of the energetic electron, $\omega_{ce}$ is the electron cyclotron frequency, and $\gamma$ is the relativity factor. For energetic electron (>100 keV) driven instability, $l = -1$ or 0 will be measured by the high-frequency magnetic diagnostic (1–180 MHz).

Unlike in runaway discharges, where the parallel temperature is much higher than the vertical temperature [27–29], the vertical temperature of energetic electrons is of the same order as the parallel temperature when ECW heats the vertical velocity of electrons [30, 31]. However, EXL-50 exhibits high-frequency electromagnetic influences (1–180 MHz) and significant frequency chirping, as predicted by the Berk-Breizman (BB) model as a result of the fully nonlinear interaction between energetic particles and wave [32–35]. The study of energetic particle instability is important for understanding the physical image of wave-particle interactions, and energetic particle confinement.

The flow of the paper is as follow: Section 2 describes the diagnostic setup and Section 3 discusses measurements, preliminary analysis, and future work.

## 2. Diagnostic setup

### 2.1. EXL-50

The major and minor radii of EXL-50's plasmas are approximately 0.58 m and 0.41 m, respectively. The toroidal magnetic field ($B_t$) in EXL-50 is approximately 1 T at r ~ 0.2 m, and the aspect ratio is approximately 1.5. Currently, EXL-50 uses three sets of ECW systems (28 GHz) to heat the plasma and drive the plasma current (Figure 1(a)). System #1 (source power of gyrotron

~50 kW) was primarily used to produce the initial plasma and form a closed flux surface. Systems #2 and #3 (source power of gyrotron ~400 kW) were used to increase the plasma current and sustain the current flattop for several seconds. Discharges with plasma currents substantially above 100 kA with a 100 kW ECW are routinely obtained, and the plasma current is mainly carried by energetic electrons [36, 37].

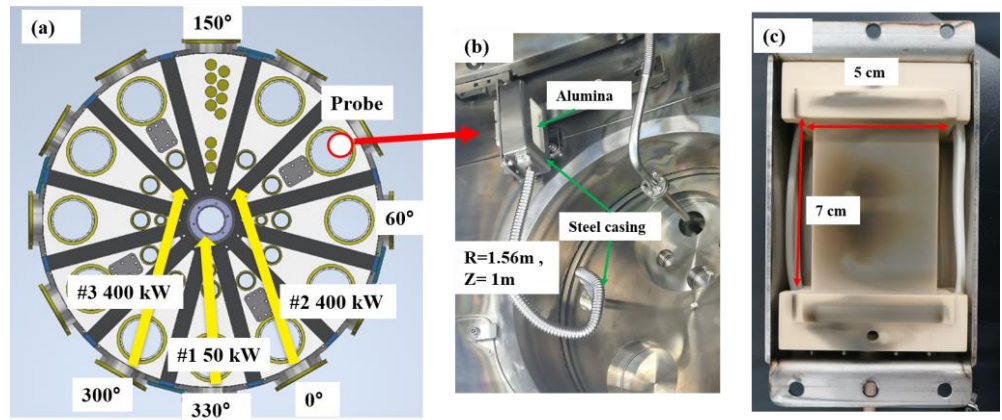

Figure 1. (a) Top-view of EXL-50 spherical torus. The magnetic loop of high-frequency magnetic probe and three sets of ECW heating systems are shown. (b) Specific location and (c) structure of the loop for toroidal magnetic fluctuation measurement. The loop is located at the top of the low field side 90° port.

## 2.2. High-frequency magnetic diagnostic

The probe of the high-frequency magnetic diagnostic is mounted on the low-field side (R~1.56 m and Z~1 m, behind the limiter ~4 cm) of the EXL-50 90° port, as shown in Figures 1(a) and 1(b). The probe, parallel to the $B_t$ direction, is a double-turn loop made of copper with a Teflon insulating sleeve (Figure 1(c)). The Teflon insulating sleeve is used to provide DC isolation that is required to protect sensitive electronics.

The length and width of the probe were 7 and 5 cm, respectively. The loop area of the probe was approximately 70 cm$^2$. A frame made of stainless steel (316 L) was created to protect the probe. The detected radio-frequency (RF) signal was transmitted to the atmosphere via a high-temperature, high-vacuum-tolerance coaxial transmission line and coaxial vacuum feedthrough.

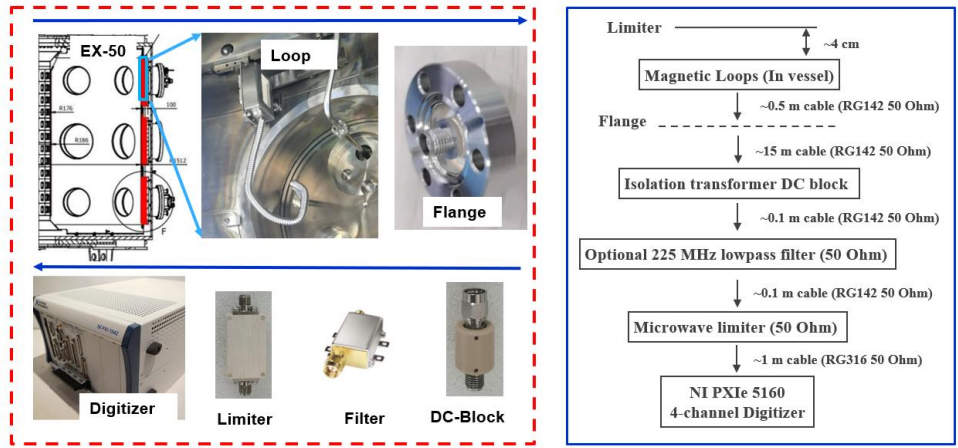

Figure 2. Hardware schematic for high-frequency magnetic diagnostic, beginning with the in-vessel loops and ending at the high-speed digitizer. All cables are 50 Ω and coaxial.

Figure 2 depicts a block diagram of high-frequency magnetic diagnostics. When high-frequency magnetic fluctuations appeared in the plasma, electrical currents were induced in the loops. The resulting voltage oscillations will initially travel through a 0.5 m coaxial vacuum cable (RG142, 50 Ω Coaxial Cable from Lair Microwave.), that can withstand a maximum baking temperature of 200 °C. The signal will then be transmitted to the airside via a cable coaxial feedthrough (IFDCF012012, SMA 50 Ω, floating shield, from Kurt J. Lesker). A 15 m cable coaxial was used to keep the electronics system (DC block, low-pass filter, RF limiter and high-speed digitizer) away from the device. The DC block (PE8212, 0.01-18 GHz, SMA 50 Ω, Breakdown Voltage 200 V, from Pasternack) was utilized to provide DC isolation when the Teflon insulating sleeve was ruptured. The low-pass filters (PE8722, DC-225 MHz, SMA 50 Ω, from Pasternack) were used to filter signals greater than 225 MHz. An RF limiter (HWLT0001300012, 0.001-3 GHz, SMA 50 Ω, HengWei Microwave) was installed to limit excessive voltage oscillations. Finally, the voltage oscillations were recorded using a high-speed digitizer. The fast analog-to digital converter digitizer (NI PXIe 5160) has a 10-bit depth, 500 MHz bandwidth, and 2.5 GS/s sampling rate. The digitizer records data in the form of time series voltage fluctuations with 50 Ω input impedance. In addition, 2 GB of data were recorded per shot for a typical discharge length of 1 s. It should be noted that we used a variety of low-pass filters (DC-30 MHz, DC-100 MHz, DC-130 MHz, DC-200 MHz, and DC-500 MHz), depending on the sampling rate to choose the filters.

Prior to installation, the probe was subjected to a laboratory test. Experiments were conducted to determine the attenuation coefficient of the probe and to demonstrate that the diagnostic system could acquire the spectrum of an RF wave. A simplified laboratory test experiment is depicted in Figure 3(a). A coupled loop with a metal frame was connected to ports 1 (source, S1) and 2 (receiving, S2) of a vector network analyzer with 50 Ω coaxial transmission line. The loop spacing is approximately 50 mm. The $S_{21}$ parameter is shown in Figure 3(b). The measurement of $S_{21}$ (black line) between the two loops revealed that the sensitivity of the loop was approximately constant in the frequency range of 20–40 MHz. In the range of 60–80 MHz, the variation in $S_{21}$ was large. The experimental measurements results (with frame) are less than the theoretically calculated value with 50 Ω input impedance (S1 as the input port and S2 as the receive port without metal frame). Because of the high intensity of EXL-50 plasma radiation, the loops are able to effectively measure the plasma radiation signal (Figure 3(c)), thereby completing the experimental investigation of energetic electron-driven instability on EXL-50.

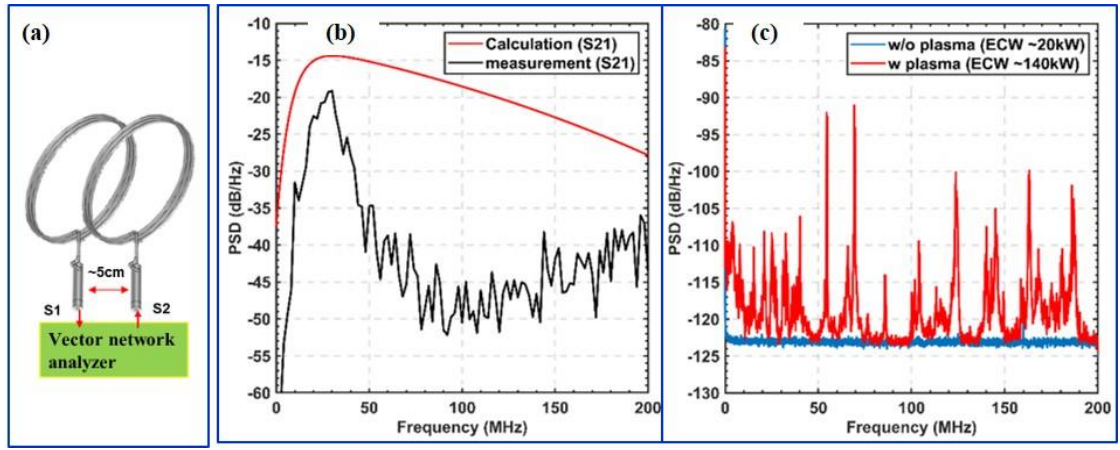

Figure 3. (a) The network analyzer (10 MHz–40 GHz) was used to get the attenuation coefficient of the probe. (b) Transfer function $S_{21}$. The black line is a measured result with a vector network analyzer (VNA) and the red line is a theoretical calculation value. (c) The power spectra of signal measured by high-frequency magnetic probe without plasma (ECW~20 kW) and with plasma (ECW~140 kW)

Figure 4 shows, from top to bottom, plasma current, hard X-ray intensity [38], and high-frequency oscillations for a typical small disruption discharge. In EXL-50, the 50 kW ECW (~20 kW injection power) is generally used to initiate breakdown plasma and plasma current from 0.0–0.2 s. Then, 400 kW ECW (~140 kW injection power) is injected to achieve high current and maintain current flattop. As depicted in Figure 4(a), the plasma current cannot increase because of disruption after the injection of high-power ECW.

With a 400 kW ECW, a significant steep drop in current was observed alongside a strong hard X-ray intensity and high-frequency signal. The acquisition system (1 MΩ, maximum input voltage 42 V) was broken by the strong radiation after 0.5 s.

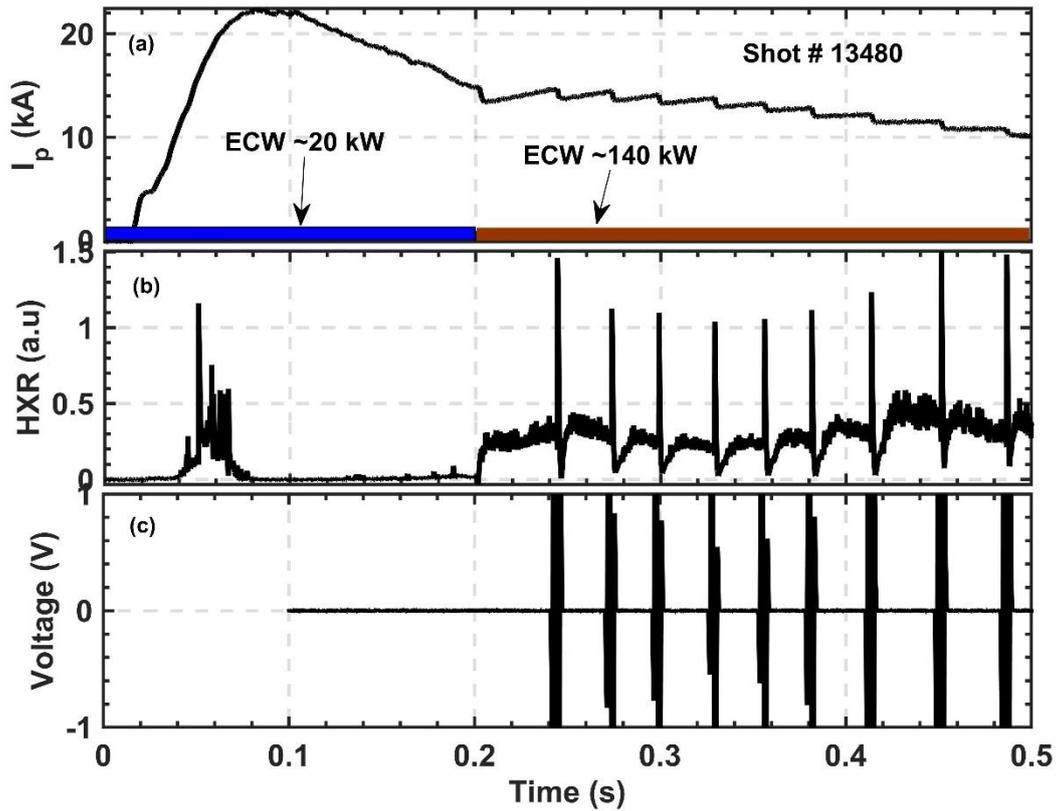

Figure 4. Temporal evolution of (a) plasma current, (b) Hard X-ray, and (c) raw trace of fluctuation.

Following this discharge, an RF limiter was utilized to protect the acquisition system. The RF limiter operating in the frequency range of 1 MHz–3 GHz is capable of handling up to 40 W of CW input power and 1,000 W of transient input power. The main parameters of the RF limiter are as follows: 40 W continuous wave input power, 1,000 W peak power, 12–18 dBm flat leakage, 2 dB max insertion loss, 1.5:1 typical input VSWR, and a fast response and short recovery time of 1 μs. It should be mentioned that data with a raw signal voltage exceeding the limiter limit voltage (~2 V) were discarded.

## 3. Experimental results

### 3.1 High-frequency (chirping) oscillation

Although the laboratory measurements showed good correspondence with the probe at 20–40 MHz,

the high-frequency oscillations of the EXL-50 device were so strong that we measured effective signals at 1–180 MHz (Figure 3(c)). Figure 5 depicts a typical steady-state plasma discharge of EXL-50, with plasma density, current, and hard X-ray intensity, from top to bottom. The ECW power was approximately 140 kW, the plasma current was ~90 kA, and the line integral density was $0.8 \times 10^{18}$ m$^{-2}$. Strong, hard X-rays were also detected when a closed magnetic flux was formed. The high-frequency magnetic diagnostic data acquisition system operates between 1.5–2 s during discharge. The spectrum from 1,680–1,698 ms is shown in Figure 6, where distinct high-frequency oscillations were observed. Oscillations between 2–4 MHz (Figure 6(b)) exhibit significant frequency chirping with multi-harmonic frequency oscillations, and oscillations at 24 MHz (Figure 6(c)) exhibit multi-harmonic frequency. These are consistent with the results of the high-energy particle-driven instability with an increased drag effect (Figure 6(b)) and large diffusion (Figure 6(c)) predicted by the BB model [32–35].

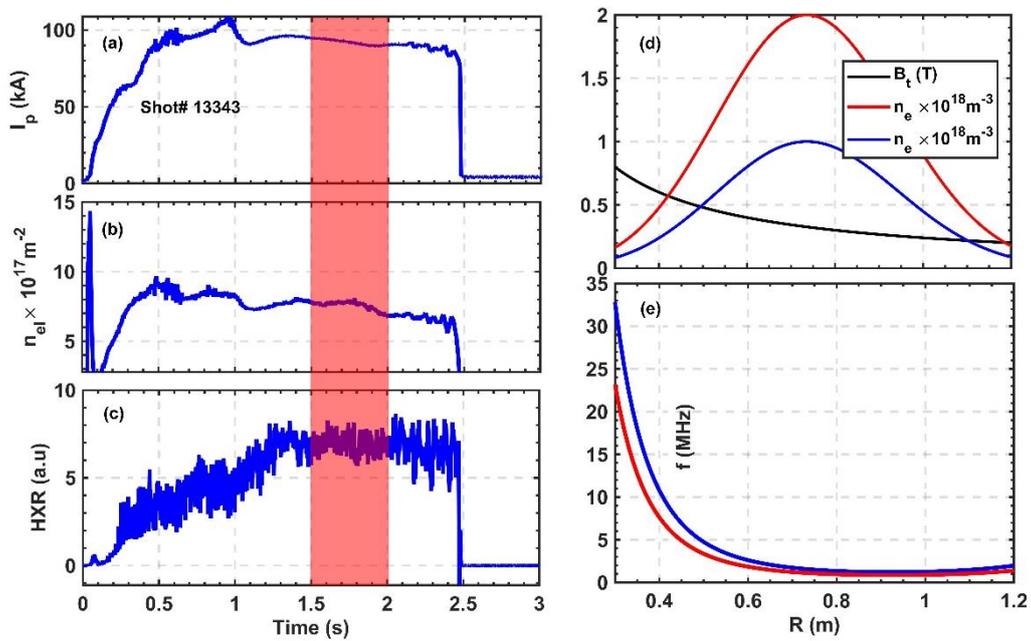

Figure 5. Temporal evolution of (a) plasma current, (b) integral density, and (c) Hard X ray; Radial profile of (d) Bt, n$_{el}$ and (e) Alfvén wave (AE) frequency.

## 3.2 Candidate instabilities

Assuming two density profiles (Figure 5 (d) an Alfvén wave (AE) frequency for s=0 (where s is the radial mode number), m=0 and n=1 ($f = v_A/2\pi R$) was assumed. As shown in Figure 5(e), on the high-field side, the AE frequencies were approximately 4–30 MHz (R<0.5 m). The AE frequencies are approximately 2–4 MHz when R > 0.5 m, which corresponds to the frequency range where significant frequency chirping was observed during the experiment. Therefore, the high-frequency electromagnetic oscillations observed were likely Alfvén waves excited by the energetic electrons (l=0 or -1).

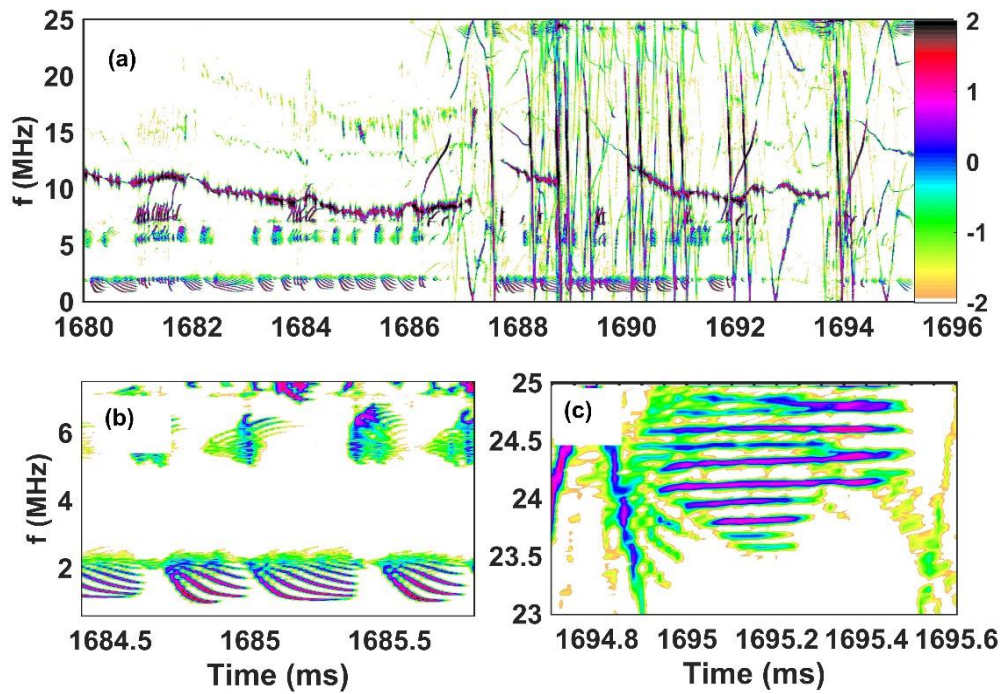

Figure 6. RF magnetic fluctuations for EXL-50 obtained from high-frequency magnetic diagnostic. Spectra during 1,680–1,696 ms (a), 1,684–1,686 ms (b), and 1,694–1,695.6 ms (c).

## 4. Conclusion

For research on high-frequency magnetic upheaval excited by energetic particles. A high-frequency magnetic probe has been designed and installed on EXL-50. An RF limiter was utilized to protect the acquisition system and prevent high-intensity radiation during steep current drops. Multiple high-frequency oscillations, most likely Alfvén waves excited by energetic electrons, were observed with the high-frequency magnetic probe. Several possible future upgrades could be made to improve and expand this high-frequency magnetic probe include improving loop (without frame), bandwidth, amplitude and spatial calibration. In future experiments in EXL-50, the plasma density and toroidal magnetic field will be scanned to investigate the dispersion relation of the oscillations and understand the effect of the oscillations on the energetic electron confinement and instability.

**Acknowledgments**

The presented research was supported in part by the National Natural Science Foundation of China under Grant No. 11705151. The authors would like to thank Yumin Wang, Xin Zhao, Guo Dong, Yubao Zhu, as well as the diagnostic group and operation teams at Energy Innovation (ENN), for their assistance in installing the diagnostics and achieving high-quality plasma discharges.

**Data availability**

The data supporting the findings of this study are available upon reasonable request from the corresponding author.